\documentclass[12pt]{article}
\usepackage{epsfig}
\usepackage{amssymb}
\usepackage{amsmath,amsfonts,amssymb,graphicx}
\newcommand{\beq}{\begin{eqnarray}}
\newcommand{\eeq}{\end{eqnarray}}
\begin{document}
\title{Improved Theory of Neutrino Oscillations in Matter}
\author{Leonard S. Kisslinger\\
Department of Physics, Carnegie Mellon University, Pittsburgh PA 15213 USA.}
\maketitle
\date{}
\begin{abstract}
 This is revision of the S-Matrix theory of neutrino oscillations used
for many years.  We evaluate the transition probability of a $\mu$
to $e$ neutrino without an approximation used for many theoretical studies,
and find important differences which could improve the extraction of neutrino
parameters from experimental data in the future. 
\end{abstract}

\noindent
.PACS Indices:11.30.Er,14.60.Lm,13.15.+g
\vspace{1mm}


\section{Introduction}

  In previous research on neutrino oscillations we used S-Matrix theory 
with three active neutrinos to evaluate time reversal violation\cite{hjk11}, 
CP violation\cite{hjkz11}, and recently\cite{khj12}  the transition 
probability $\mathcal{P}(\nu_\mu \rightarrow$ $\nu_e$) . One of the main 
objectives of Ref.\cite{khj12} was to calculate neutrino oscillations
and anti-electron neutrino disappearance to help in the measurement of 
$\theta_{13}$ from Daya Bay, Double Chooze, and RENO data. For this it is
essential to have an accurate theory of neutrino transition probabilities.

  Our previous research using S-Matrix theory with a 3$\times$3 mixing matrix 
for three active neutrinos was based on earlier 
publications\cite{as97,f01,ahlo01,jo04}. 
In carrying out the evaluation of $\mathcal{P}(\nu_\mu \rightarrow$ $\nu_e)$
we used an approximation for an important function, $I_{\alpha^*}$, found in this
earlier work, such as Ref\cite{ahlo01}, whose formulation we used. 
In the present work we evalute  $I_{\alpha^*}$ exactly and find corrections to 
 $\mathcal{P}(\nu_\mu \rightarrow$ $\nu_e$). Our objective is to improve
$\mathcal{P}(\nu_\mu \rightarrow$ $\nu_e$), and thereby the extraction of
important parameters from neutrino oscilation data.

\section{$\mathcal{P}(\nu_\mu \rightarrow$ $\nu_e$) Derived Using 
S-Matrix Theory}

Neutrinos are produced as $\nu_f$, with $f$ =flavor=$e, \mu, \tau$. 
Neutrinos with definite mass are $\nu_m$, m=1,2,3. The flavor neutrinos
are related to the neutrinos with definite mass by the 3$\times$3 unitary 
matrix, $U$,
\beq
\label{f-mrelation}
      \nu_f &=& U\nu_m \; ,
\eeq
where $\nu_f, \nu_m$ are 3$\times$1 column vectors and $U$ is 
($sin\theta_{ij} \equiv s_{ij}$, etc). 
\newpage

$U$=
$\left( \begin{array}{clcr} c_{12}c_{13} & s_{12}c_{13} &s_{13} e^{-i\delta_{CP}}\\
-s_{12}c_{23}-c_{12}s_{23}s_{13}e^{i\delta_{CP}} &c_{12}c_{23}-
s_{12}s_{23}s_{13}e^{i\delta_{CP}}
& s_{23}c_{13} \\ s_{12}s_{23}-c_{12}s_{23}s_{13}e^{i\delta_{CP}} & -c_{12}s_{23}
-s_{12}c_{23}s_{13}e^{i\delta_{CP}} &  c_{23}c_{13} \end{array} \right)$
\vspace{3mm}

We use $c_{12}=.83,\;s_{12}=.56,\;s_{23}=c_{23}=.7071$, $s_{13}$= 0.19, and
$\delta_{CP}$=0. 

The transition probability $\mathcal{P}(\nu_\mu \rightarrow$ $\nu_e$) is 
obtained from the S-Matrix element $S_{12}$, where  $i\frac{d}{dt}S(t,t_0) = 
H(t) S(t,t_0)$, with $H(t)$ the Hamiltonian:
\beq
\label{PmueS12}
 \mathcal{P}(\nu_\mu \rightarrow \nu_e) &=& (Re[S_{12}])^2 + (Im[S_{12}])^2
\; .
\eeq
\vspace{3mm}

From Ref\cite{khj12}
\beq
\label{ReImS12}
    Re[S12]&=&s_{23}a[cos(\bar{\Delta}L)Im[I_{\alpha^*}]-sin(\bar{\Delta}L)
Re[I_{\alpha^*}]] \nonumber \\
    Im[S12]&=&-c_{23}sin2\theta sin\omega L-s_{23}a[cos(\bar{\Delta}L)
Re[I_{\alpha^*}] \nonumber \\
   &&+sin(\bar{\Delta}L)Im[I_{\alpha^*}] \; ,
\eeq
with
\beq
\label{Deltadelta}
    \bar{\Delta}&=& \Delta-(V+\delta)/2 \nonumber \\
    \Delta&=&\delta m_{13}^2/(2E)  \\
     \delta&=&\delta m_{12}^2/(2E) \nonumber \; ,
\eeq
where the neutrino mass differences are 
$\delta m_{12}^2=7.6x10^{-5}(eV)^2$, $\delta m_{13}^2=2.4x10^{-3}(eV)^2$, 
$sin2\theta= s_{12}c_{12} \frac{\delta}{\omega}$, $a= s_{13}(\Delta-s_{12}^2 
\delta)$, and $E$ is the neutrino energy. Note that $t\rightarrow L$, 
where $L$ is the baseline, for $v_\nu \simeq c$. The neutrino-matter potential
$V=1.13\times 10^{-13}$ eV. 

The main quantity of interest in our present work, $I_{\alpha^*}$, is
\beq
\label{Ialpha*}
    I_{\alpha^*}&=& \int_{0}^{t} dt' \alpha^*(t')e^{-i\bar{\Delta}t'} \; ,
\eeq
with $\alpha(t)= cos(\omega t)-icos2\theta sin(\omega t)$, $\omega=
 \sqrt{\delta^2 +V^2-2\delta V cos(2\theta_{12})}/2$.

In Ref.\cite{khj12}, as in Ref.\cite{ahlo01}, one used 
$\delta,\omega \ll \Delta$ to obtain
\beq
\label{ReImI}
     Re[I_{\alpha^*}]&\simeq& sin\bar{\Delta}L/\bar{\Delta} \nonumber \\
    Im[I_{\alpha^*}]&\simeq& (1-cos\bar{\Delta}L)/\bar{\Delta} \; .
\eeq

On the other hand if one uses Eq(\ref{Ialpha*}) to evaluate 
$I_{\alpha^*}$, one obtains
\beq
\label{ReImIexact}
     Re[I_{\alpha^*}]&=&  [(\omega-\bar{\Delta}cos2\theta)cos\bar{\Delta} L
sin\omega L \nonumber \\
   &&-(\bar{\Delta}-\omega cos2\theta)sin\bar{\Delta} Lcos\omega L]/
(\omega^2-\bar{\Delta}^2) \\
Im[I_{\alpha^*}]&=& [\bar{\Delta}+\omega cos2\theta-
(\bar{\Delta}+\omega cos2\theta)cos\bar{\Delta} Lcos\omega L \nonumber \\
  &&-(\omega+\bar{\Delta}cos2\theta)sin\bar{\Delta} Lsin\omega L]/(\omega^2
-\bar{\Delta}^2).
\eeq

From this we obtain an improved $\mathcal{P}(\nu_\mu \rightarrow$ $\nu_{e}$)=
 (Re[$S_{12}])^2$ +(Im[$S_{12}])^2$.
\clearpage

\hspace{3cm}$\mathcal{P}(\nu_\mu \rightarrow$ $\nu_{e}$) with old and precise 
$I_{\alpha^*}(E,L)$.
\vspace{6cm}

\begin{figure}[ht]
\begin{center}
\epsfig{file=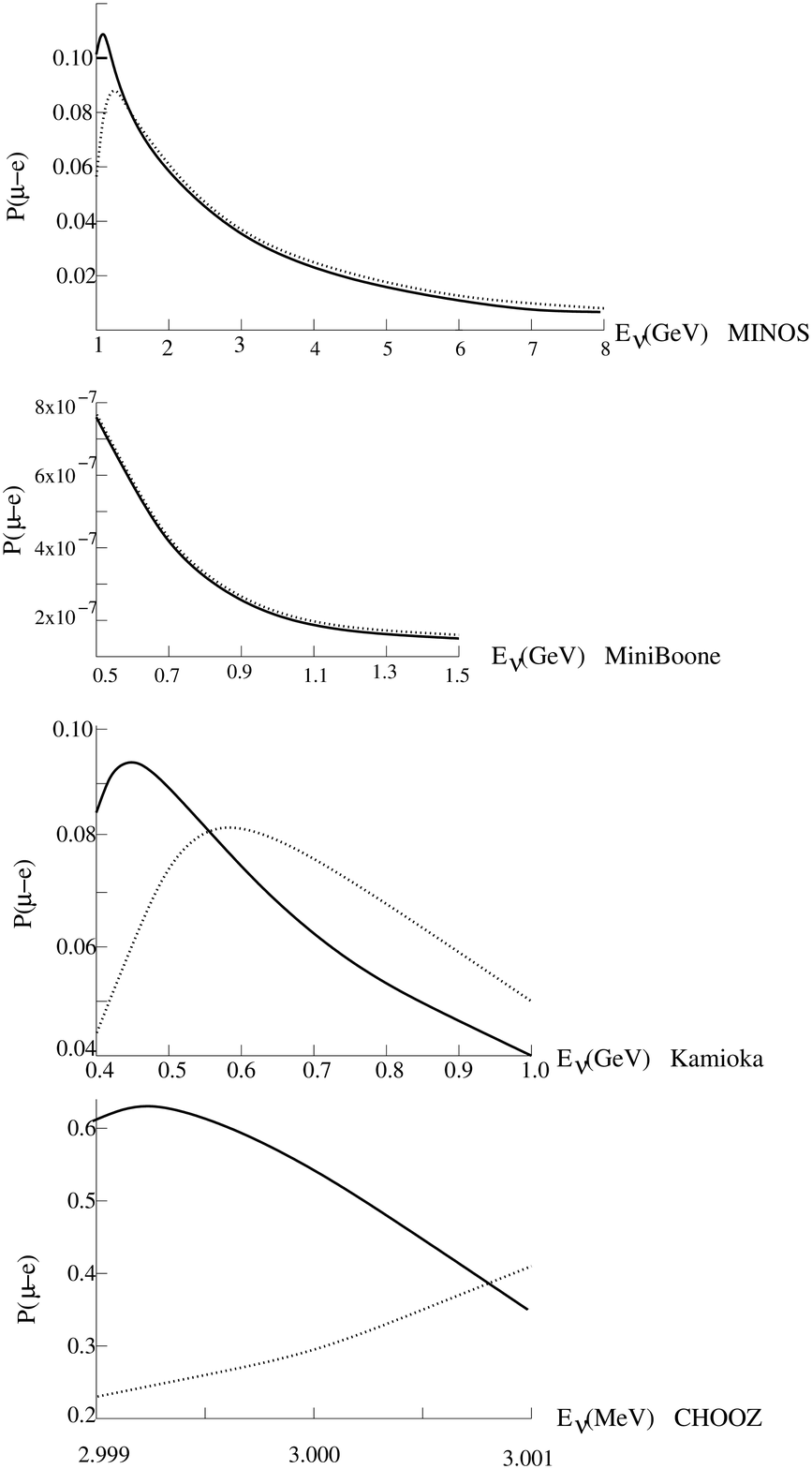,height=12cm,width=10cm}
\end{center}
\caption{$\mathcal{P}(\nu_\mu \rightarrow\nu_e)$ 
for MINOS(L=735 km), MiniBooNE(L=500m), JHF-Kamioka(L=295 km), and 
CHOOZ(L=1.03 km) using the improved 3$\times$3 mixing matrix. Solid curve for 
precise  $I_{\alpha^*}(E,L)$ and dashed curve for approximate  $I_{\alpha^*}(E,L)$.
$s_{13}$=0.19} 
\end{figure}
\newpage

\section{Conclusions}

As one can see from the figure, for a large baseline (MINOS) or energy 
(MiniBooNE) the corrections are small. For some baselines and energies,
including JHF-Kamioka and CHOOZ, the 
corrections to the neutrino transition probability $\mathcal{P}(\nu_\mu 
\rightarrow$ $\nu_{e}$)  are large. This will 
modify the predictions of CP and T violation, and could have a large effect
on the extraction of parameters via neutrino oscillations, such as  $s_{13}$ 
from $\mathcal{P}$($\bar{\nu}_e$$ \rightarrow$$\bar{\nu}_e$) by 
Daya Bay\cite{DB3-7-12}, Double Chooz\cite{DC11}, and, RENO\cite{RENO10}.
This is an important result in neutrino physics, and the precise
expression for $I_{\alpha^*}$ we have derived should be used for all future
studies of neutrino oscillations.

\Large
{\bf Acknowledgements}
\vspace{3mm}

\normalsize
This work was carried out while LSK was a visitor at Los Alamos
National Laboratory, Group P25. The author thanks Dr. William Louis for 
information about recent and future neutrino oscillation experiments,
and Dr. Mikkel B. Johnson for many helpful discussions.

\end{document}